\documentclass{amsart}

\usepackage[T1]{fontenc}

\usepackage{amsmath,amsfonts,amssymb}
\usepackage{xspace}
\usepackage{graphicx}
\usepackage{array}
\usepackage[table,dvipsnames]{xcolor}

\usepackage{siunitx}
\usepackage{enumitem}
\usepackage{epstopdf}
\usepackage{tikz}
\usepackage{chemformula}
\usepackage{hyperref}
\hypersetup{
    colorlinks=true,
    linkcolor=blue,
    filecolor=magenta,      
    urlcolor=cyan,
}

\newcommand{\epsi}{\ensuremath{\varepsilon}}
\newcommand{\fracpar}[2]{\frac{\partial #1}{\partial #2}}

\newcommand{\Real}{\mathbb{R}}

\newcommand{\abs}[1]{\left|#1\right|}

\newcommand{\grad}{\nabla}
\newcommand{\dive}{\nabla\cdot}

\DeclareMathOperator{\rank}{rank}

\usepackage{soul}
\soulregister\cite7
\soulregister\ref7
\soulregister\eqref7
\soulregister\num7
\soulregister\pageref7

\setul{0mm}{1mm}
\setstcolor{red}
\sethlcolor{yellow}

\def\mytilde{\hspace*{-2pt}
  \begin{tikzpicture}
    \draw[red] (0, 0) .. controls (1pt, 1pt) .. (2pt, 0) .. controls +(1pt, -1pt) .. +(2pt, 0);
  \end{tikzpicture}
}
\newbox\mybox
\setbox\mybox\hbox{\mytilde}
\makeatletter
\def\xavierul #1{\begingroup\def\SOUL@ulleaders{\leaders\hbox{\raisebox{-4pt}[0pt][0pt]{\hbox to 5pt {\copy\mybox}}}\relax}\ul{#1}\endgroup}
\makeatother
\colorlet{comcolor}{blue!80!yellow!20!white}
\colorlet{softblue}{blue!20!yellow!30!white}

\title{Robust chemical solver for fully-implicit simulations}
\author{Colin McNeece \and Xavier Raynaud \and Halvor Nilsen  \and Marc Hesse }
\date{}

\DeclareSIUnit\equiv{eq}

\begin{document}

\maketitle

\thispagestyle{empty}

\section*{Abstract}

The study of geological systems requires the solution of complex geochemical relations. We present an implementation of a chemical solver which can handle various types of models, including surface chemistry. The implementation is done in view of easy coupling with flow simulations to obtain a fully-coupled, fully-implicit solver for chemical reaction transport equations applicable to realistic reservoir models.

\section{Introduction} 

In this paper, we present simulation and implementations strategies for computing
solutions of a variety of chemical models. The solver is open-source and available
for download \cite{MATCH}. The implementation supports non-isothermal
multicomponent aqueous complexation, surface complexation (including the CD-MUSIC
model), ion exchange, and dissolution/precipitation. After rewriting the standard
chemical equilibrium in a generic form, we explain how a standard log-log
transformation of the residuals and primary variables can improve the robustness
of the Newton method. We derive simple apriori bounds on the system unknowns and
show numerically how they contribute to improve the convergence. Our solving
strategy includes the computation of good initial guesses calculated explicitly by
sequentially adding model complexity.

The implementation relies heavily on automatic differentiation which eliminates
the burden of computing analytically and implementing the jacobians of the
residuals. We also take profit of an existing platform to assemble the equations
and run the Newton steps. We use open-source MRST and refer to
\cite{krogstad2015mrst} for detailed explanation on how the tool can be used in
this setting. More significantly, MRST offers supports for unstructured grids
which are typically associated with realistic reservoir models. Discrete
differentiation operators for finite volume methods are readily available. The
discrete differential operators and automatic differentiation are the two
ingredients which significantly simplify the implementation of fully-implicit
multi-physics solvers for reservoir flow simulations. Prototyping allows for
testing of solver strategies and, once the best methods are identified,
computational speed can be optimized, possibly by switching to a compiled
language.

The chemical solver has been validated for each model against PHREEQC
\cite{parkhurst2013description} which is the reference chemical solver for
applications in geochemistry. The benchmark tests are not included in this paper
but are available in the manual which can be found in \cite{MATCH}.

The coupled chemistry-transport equations are presented here and the simulation of
these equations, using fully-coupled and fully-implicit methods, is clearly one of
the main motivations behind the implementation of the chemical solver we have
presenting here. Still, at the moment of publication, we only have one-sided
coupled simulations running, that is, simulations where the chemical composition
has no effect on the flow. We are working on a fully coupled test, in our case, a
carbon system which includes the dissolution of calcite and thus the increase of
permeability. We plan to release results in a very near future. Different
numerical schemes (for example splitting strategies) and additional physics can
also be incorporated into the existing tools through the object-oriented framework
employed by MRST.

Identifying the generic structure of chemical equations allows us to develop a
flexible user interface. The solver can handle arbitrarily complex geochemical
systems with any choice of species or element concentration as input depending on
what is known of the chemical system. The user interface is essentially based on
parsing of input strings and the string inputs use standard notations in
chemistry.

\section{Modeling equations}
\subsection{General structure of the equilibrium equations}

In this section, we present a brief derivation of the equilibrium equations with
an emphasis on the algebraic structure of the equations.  We remark that the
system of equations split into two sets: A first set of equations which are linear
in the logarithm of the concentrations and a second set where the equations are
linear with respect to the concentration. In the following section, this structure
is used to justify the robustness of the Newton solver when using a log-log
transformation.

Let us consider $n_c$ chemical species, $\{C_i\}_{i=1}^{n_c}$, in quantities
$\{N_i\}_{i=1}^{n_c}$, entering into $n_e$ chemical reactions.  The stochiometric
relations can be written (by moving all terms to the left side, for notational
convenience):
\begin{equation}
  \label{eq:chemreac}
  a_{k,1}C_1 + a_{k,2}C_2 + \ldots + a_{k,n_c}C_{n_c} \ch{<>[ $K_k^+$ ][ $K_k^-$ ] 0},
\end{equation}
where $a_{k,i}$ are integers which can also be negative numbers. Let us consider a
variation $\delta N$ in the quantities of components. This variation is a result
of the chemical reactions. Let $\nu_k$ be the number of times the reaction $k$ has
occurred.  We have
\begin{equation}
  \label{eq:reacnum}
  \delta N = -\sum_{k=1}^{n_e}\nu_k a_k = -A^t \nu,
\end{equation}
where $a_k$ denotes the row of $A$ and $\nu = (\nu_1, \ldots, \nu_{n_e})$. Hence,
$\delta N \in \rank(A^t)$, that is, $\delta N \in \ker(A)^{\perp}$. Let us $V$
denotes a basis of $\ker(A)$ and $W$ a basis of its complement. Provided that
there are no redundancies in the chemical reactions described by $A$, its rows
will be linearly independent, and we can (for instance) choose $W = A^t$. From
\eqref{eq:reacnum}, we have
\begin{equation}
  \label{eq:conseq}
  V^t\delta N = 0,
\end{equation}
which shows that the kernel of $A^t$ identifies \textit{natural} linear
combinations that are preserved at equilibrium. To see this more clearly, let us
consider the kinetic equations, before the equilibrium is reached. For every
chemical reaction, there corresponds the kinetic equation given by
\begin{equation}
  \label{eq:kineq}
  \frac{dN_i}{dt} = F_i,
\end{equation}
where $F_i$ is defined as
\begin{equation}
  \label{eq:defFi}
  F_i = -\sum_{k=1}^{n_e} a_{k,i}G_k = -A^tG
\end{equation}
for 
\begin{equation}
  \label{eq:defGk}
  G_k = k_k^+\prod_{a_{k,i> 0 }}N_i^{|a_{k,i}|} - k_k^-\prod_{a_{k,i<0}}N_i^{|a_{k,i}|}
\end{equation}
and $G=(G_1,\ldots, G_{n_e})$. The definition of $G_k$ is valid if all components
which enter the chemical equation $k$ have activities equal to their
concentration, which we assume for now (it is not always true, e.g. for $\ch{H2O}$
or for precipitated substances such as calcite). Let
\begin{equation}
  \label{eq:defMN}
  M = V^tN\quad\text{ and }\quad P = W^tN.
\end{equation}
The governing equations \eqref{eq:kineq} can then be decomposed in two sets such that
\begin{equation}
  \label{eq:twosets}
  \frac{d}{dt} M = 0\quad\text{ and }\quad \frac{d}{dt}P = W^tF.
\end{equation}
This decomposition shows that the linear combinations given by the kernel basis
$V$ are invariant quantities. We do not need such mathematical approach to
establish the existence of invariant quantities. Indeed, in the system we are
considering, the chemical species are molecules, meaning that they correspond to a
given combination of atoms. The atoms are constitutive elements and the total
number of each type of atom is preserved. The concentration of a given atom is
obtained as a linear combination of the concentration, weighted by the occurrence
number of the atom in the molecules. Note that, in the presentation above, we did
not mention atoms and only the species, see \eqref{eq:chemreac}. Thus, we may
wonder how, ignoring the existence of atoms, we managed to infer the existence of
invariants, which in turn correspond to the atoms. This is explained by the fact
that the chemical reaction equations are always well-balanced in the sense that
each of them, when written explicitly, have preserved quantities, which precisely
correspond to the atoms involved in the composition of the molecules entering the
equation. We recover these invariants, in a purely algebraic manner, by looking at
the kernel of the reaction matrix $A$. However, the atomic nature of the species
gives us a precious additional structure which turns to be determinant in the
solution procedure we will describe later. In the previous algebraic derivation,
the matrix $V$ is not unique and we have no result on the sign of the coefficients
of $V$. But now, by using the distribution of each atom into the species, we know
that we can find a matrix $V$ such that all the coefficients of $V$ are
\emph{non-negative}, see \eqref{eq:Vccmat} in the illustrative example below.
There exist a well-known invariant of the reaction equations: Charge balance. The
charge balance, as the conservation of atom type, is also explicitly enforced in
every reaction equation. However the decomposition of the charge in terms of the
species does not lead to a linear combinations with only non-negative elements,
see \eqref{eq:chargedec} in the example.

From \eqref{eq:twosets}, we infer that the chemical equilibrium equation are given
by $W^tF = 0$, that is,
\begin{equation}
  \label{eq:chemeq}
  W^tA^t G  = 0.
\end{equation}
If we can choose $W=A^t$, this equation reduces to $AA^tG=0$. Since $W^tA^t$ is
invertible, and from \eqref{eq:chemeq}, we recover as expected that the chemical
equilibrium equations are given by $G=0$.

Let us simplify the notations and denote by $x\in\Real^{n_c}$ the vector of
species concentrations (denoted previously $N$) and $\hat x\in\Real^{n_c}$ the
vector of the logarithm of the concentrations, that is $\hat x_i = \ln(x_i)$. At
chemical equilibrium, we have $G=0$, which is equivalent to
\begin{equation}
  \label{eq:chemeqprod}
  \prod_{j = 1}^{n_c}x_j^{a_{k,j}} = K_k
\end{equation}
for $K_k = \frac{K_k^+}{K_k^-}$. Equation \eqref{eq:chemeqprod} is equivalent to
\begin{equation}
  \label{eq:chemeqprodlog}
  \sum_{j = 1}^{n_c}a_{k,j}\hat x_j = \hat{K}_k,
\end{equation}
where $\hat K_k = \ln(K_k)$. We introduce the concentration $X_i$ of each atom or
constitutive element. By looking at the occurrence of a given atom in each
species, we can assemble a matrix $V$ such that
\begin{equation}
  \label{eq:massconsx}
  \sum V_{j,i}x_j = X_i. 
\end{equation}
The equilibrium equations consist of \eqref{eq:chemeqprodlog} and
\eqref{eq:massconsx}, which we can rewrite synthetically as
\begin{subequations}
  \label{eq:chemeqnice}
\begin{align}
    \label{eq:chemeqnice1}
    A\hat x &= \hat K,\\
    \label{eq:chemeqnice2}
    V^tx &= X.
\end{align}
\end{subequations}
Thus, the chemical equilibrium equations consist of a set of linear equations of
either the concentrations or the logarithm of the concentration. Moreover, the set
of equations that corresponds to the linear combinations of the concentrations
only involve positive coefficients.

\subsection{Illustrative example of a carbon system}

We consider a carbon system. The reactions are the following
\begin{subequations}
  \label{eq:reactions}
  \begin{align}
    &\ch{H2O <> H^{+} + OH^{-}},\\
    &\ch{H2CO3 <> H^+ + HCO3^{-} },\\
    &\ch{HCO3^{-} <> H^+ + CO3^{2-} }.
  \end{align}
\end{subequations}
There are $n_e=3$ reactions and $n_c=9$ components, which we order as follows,
\begin{center} 
  \ch{H2O}, \ch{H^+}, \ch{OH^-}, \ch{H2CO3}, \ch{HCO3^-}, \ch{CO3^{2-}}.
\end{center}
The matrix $A$, as defined in the previous section, is given by
\begin{equation}
  \label{eq:Accmat}
  A=
  \begin{pmatrix}
    1&-1&-1& 0& 0& 0\\
    0&-1& 0& 1&-1& 0\\
    0&-1& 0& 0& 1&-1
  \end{pmatrix}.
\end{equation}
The rank of $A$ is 3. We can obtain a basis of $\ker(A)$ by considering the total
conservation of each constitutive element, which are in this case \ch{C}, \ch{O}, \ch{H}. The total conservation of \ch{C} gives
\begin{subequations}
  \label{eq:conseqCc}
  \begin{equation}
    \delta(\ch{[H2CO3] + [HCO3^{-}] + [CO3^{2-}]}) = 0,
  \end{equation}
  the total conservation of \ch{O} gives
  \begin{equation}
    \delta(m(\ch{H2O}) + \ch{[OH^-]} + 3\ch{[H2CO3]}  + 3\ch{[HCO3^-]} + 3\ch{[CO3^{2-}]}) = 0 
  \end{equation}
  and the total conservation of \ch{H} gives
  \begin{equation}
    \delta(2m(\ch{H2O}) + \ch{[H^+]} + \ch{[OH^-]} + 2\ch{[H2CO3]} + \ch{[HCO3^-]}) = 0.
  \end{equation}
\end{subequations}
Thus, we obtain the matrix
\begin{equation}
  \label{eq:Vccmat}
  V =
  \begin{pmatrix}
    0 & 1 & 2\\
    0 & 0 & 1\\
    0 & 1 & 1\\
    1 & 3 & 2\\
    1 & 3 & 1\\
    1 & 3 & 0
  \end{pmatrix},
\end{equation}
and we can check directly that $AV = 0$. Moreover, all the coefficients of $V$ are
positive, which is expected as the matrix was assembled using the conservation of
the constitutive elements entering the reactions. In comparison, the linear
combination for the conservation of charge given by
\begin{equation*}
  \delta(\ch{[H^+]} - \ch{[OH^-]} - \ch{[HCO3^-]} - 2\ch{[CO3^{2-}]}) = 0,
\end{equation*}
does not have this property. Note that it can be recovered as a linear
combination of  the columns of $V$, 
\begin{equation}
  \label{eq:chargedec}
  e = \begin{pmatrix}
    $\begin{tabular}[c]{r}
    0  \\
    1  \\
    -1 \\
    0  \\
    -1 \\
    -2 
    \end{tabular}$
  \end{pmatrix} = 4V_1 - 2V_2 - V_3.
\end{equation}

\subsection{Activity of Aqueous Species}

We relax the assumption of an ideal solution, and implement activities for all
aqueous species. We denote by $\gamma_j$ the activity of the species index by
$j$. When we consider activities, the equilibrium equation \eqref{eq:chemeqprod}
are replaced by
\begin{equation}
  \label{eq:chemeqprogact}
    \prod_{j = 1}^{n_c}(x_j\gamma_j)^{a_{k,j}} = K_k.
\end{equation}
After applying the logarithm, we get
\begin{equation}
  \label{eq:chemeqsumact}
  \sum_{j = 1}^{n_c}a_{k,j} (\hat x_j + \hat\gamma_j) = \hat K_k,
\end{equation}
where $\hat\gamma_j = \ln(\gamma_j)$. We can rewrite \eqref{eq:chemeqsumact} in a
condensed form, similar to \eqref{eq:chemeqnice1},
\begin{equation}
  \label{eq:chemeqact}
  A\hat x + A\hat \gamma = \hat K.
\end{equation}
The activity of aqueous species is determined by the extended Davies equation
\begin{equation}
  \label{eq:aqactivity}
  \log_{10}\left(\gamma_j\right) = A z_i^2\left(\dfrac{I^{1/2}}{1+I^{1/2}} - 0.3I\right),
\end{equation}
where $z_j$ is the charge of species $j$, and $I$ is the ionic strength of the bulk solution. The parameter $A$ being determined by
\begin{equation*}
    A = 1.82\times10^6 \left(e_wT\right)^{-3/2}.
\end{equation*}
where $e_w$ is the relative permeability of water, and $T$ is the temperature of the bulk solution (K) \cite{Davies1962}{}. The dielectric constant of water is determined by the polynomial function \cite{malmberg1956dielectric}{}
\begin{equation*}
    e_w = 87.740 - 0.4008(T-273.15) + 9.398\times10^{-4}(T-273.15)^2 - 1.410\times10^{-6}(T-273.15)^3.
\end{equation*}
The ionic strength of the solution is calculated by
\begin{equation}
  \label{eq:ionicstrengthdef}
  I = \dfrac{1}{2}\sum_{j=1}^{n_c} z_j C_j \delta_j
\end{equation}
where, in this case, $\delta$ removes the contribution of charge from surface
species and the electron \ch{e-}, that is $\delta_i=0$ if species $i$ is on a
surface or is \ch{e-} and $\delta_i=1$ otherwise.
  
\subsection{Surface chemistry}

Many natural aqueous systems are composed of a liquid water phase in contact with
a solid phase, such as a groundwater aquifer and the material that compose the
subsurface. As the solid is composed of repeating chemical structures, the
boundaries of the solid mark the location of under-coordinated atoms. These
truncations in the crystal structure lead to charge accumulation at the interface
of the solid and liquid phases. To balance this charge, ions and polar molecules
in the liquid water phase migrate to, and interact with the solid surface. These
chemical interactions can be long lived, forming covalent bonds, eventually
leading to the precipitation of a further solid phase. Or, can be short lived,
transient reactions, being driven by van der Waal type forces. In the present
study, we model the later and will refer to them as sorption reactions.

In general all surface chemistry models define a surface as a finite number of
sites with which chemical species in the liquid may interact. Depending on the
model, the surfaces can be either occupied or free. However, in the case of the
ion exchange model, a surface site must always be occupied; the sorption of one
ion from the liquid necessitating the release of a sorbed ion. As with chemical
reactions in the liquid, surface reactions are governed by laws of mass action,
and conservation of surface sites. Practically, the main
characteristic which mathematically distinguish surface reactions from aqueous reactions are the equations of state which govern the
activity of surface species. Many mathematical models have been developed to capture
this phenomenon both empirically, and mechanistically. The solver developed here
is capable of solving the most common model forms, which are detailed below.

\subsubsection{Activity Coefficients of Surface Species}
The activity coefficients of surface species depends on the surface chemistry model that is employed. 

The activity coefficients of species associated with ion exchange and Langmuir
type surfaces is unity, and their concentrations are determined by their active
fraction consistent with the Gaines-Thomas convention \cite{gaines1953}{}.

A more mechanistic approach to surface chemistry modeling takes the form of the so
called surface complexation models. Such models represent the mineral-liquid
interface as capacitors in series, the interfaces of those layers being planes of charge (ion) accumulation. The activity coefficients of species associated with an electrostatic surface (such as the triple layer and constant capacitance models) are determined by the potential 
and charge of the planes which sorbing the species occupy,
\begin{equation}
  \label{eq:surfAct}
    \gamma_j = \exp\left(\dfrac{F\sum_{p=1}^{n_p}\zeta_{j,o,p}\Psi_{o,p}}{RT}\right),
\end{equation}
where $F$ is Faraday's constant, $\zeta_{j,o,p}$ is the charge contribution of
species $j$ to plane $p$ of electrostatic surface $o$, $\Psi_{o,p}$, is the
electric potential of the $p^\text{th}$ plane of the $o^\text{th}$ electrostatic
surface and $R$ is the ideal gas constant \cite{Hiemstra1989a}{}. Note that a
surface species can be associated with multiple surface functional groups, but
only one electrostatic surface.

\subsubsection{Electrostatics of the surface}

The formulation of the activity coefficient for electrostatic surfaces can be
generalized as above, however, the determination of the charge and potential of
the surface depends on the electrostatic model being employed. For a full
description and comparison of different surface chemistry models consult
\cite{westall1980}, which is the text from which the constitutive relationships
implemented here are pulled.

The simplest model is the \textit{constant capacitance model}. The mineral surface
is the only layer in this model. The charge of the mineral surface, $\sigma$, is
calculated as the linear combination of charged species which reside on the
surface
\begin{equation*}
    \sigma_{o,p=1} = \dfrac{F}{S_o a_o}\sum_{j=1}^{n_c}c_j\zeta_{j,o,p=1}
\end{equation*}
where $\sigma_{o,p=1}$ is the charge density of the mineral surface of the
$o^\text{th}$ electrostatic surface (which is a constant capacitance surface), and
$S_o$ and $a_o$ are the specific surface area and slurry density of electrostatic
surface $o$. Note the charge contribution of a species to an electrostatic surface
on which it does not reside will be zero.

The constant capacitance model simulates the mineral-liquid interface as a
capacitor. The potential is therefore determined by the capacitance density of the
interface

\begin{equation*}
    \Psi_{o,p=1} = \dfrac{\sigma_{o,p=1}}{C_{o,q=1}}
\end{equation*}
where $C_{o,q=1}$ is the capacitance density of the $q^\text{th}$ layer of the
$o^\text{th}$ electrostatic surface. Note that charge neutrality is not possible
for a constant capacitance surface.

The \textit{triple layer model} is a more accurate model which approximates the
mineral-liquid interface as three capacitors in series. The first two, starting
from the mineral surface have a constant capacitance density, the outer most layer
has a variable capacitance density as determined by the properties of the bulk
solution, according to the Grahame equation. Just as in the constant capacitance
model the charge of a plane is the linear summation of charged species that reside
on the plane, while the charge density of the outer layer is determined by the
Grahame equation. We have
\begin{subequations}
  \label{eq:charge}
  \begin{equation}
    \label{eq:charge1}
    \sigma_{o,p} = \dfrac{F}{S_o a_o}\sum_{j=1}^{n_c}c_j\zeta_{j,o,p}
  \end{equation}
  for $p=1,2$ and
  \begin{equation}
    \label{eq:charge2}
    \sigma_{o,3} = - (8\times10^3 R T I e_o e_w)^{1/2}\sinh\left(\dfrac{F\Psi_{o,3}}{2RT}\right)
  \end{equation}
\end{subequations}
The charge-potential relationship for the triple layer surface is then determined by
\begin{align}
\sigma_{o,p=1}  &= C_{o,q=1}\left(\Psi_{o,p=1} -\Psi_{o,p=2}\right)\label{eq:pot1}\\
\sigma_{o,p=3}  &= C_{o,q=2}\left(\Psi_{o,p=3} -\Psi_{o,p=2}\right)\label{eq:pot2}
\end{align}
Finally the triple layer surface must be charge neutral
\begin{equation}
    0 = \sum_{p=1}^{n_p} \sigma_{o,p}.\label{eq:neutral}
\end{equation}

\subsubsection{Basic stern model and diffuse layer model}
These two models are limiting cases of the triple layer model. The Basic stern
model is achieved when there is no sorption on the $p=2$ plane , and thus the
potential drop across the $q=2$ layer is negligible. This is approximated by
disallowing sorption reactions to occur on the $p=2$ plane and setting the
capacitance density of the $q=2$ layer to a very large number (say 1000
coulombs/m$^2$). The diffuse layer model approximates the mineral-liquid interface
as simply the diffuse ion cloud. This is achieved by ignoring the two inner
constant capacitance layers of the triple layer by disallowing sorption on the $p=2$
plane, and setting the capacitance density of the inner two layers to a large
number as before.

\subsection{Practical Implementation}
Given conventions within the geochemical community, there is some subtlety within
the above formulation which requires detailing.

Primary among these is local chemical equilibrium, an assumption that is
implemented in the geochemical code. Such an assumption is largely accepted for
aqueous reaction. However, the formation of solids is a time dependent problem. In addition, formation/removal of solid and other phases requires a volume change. Thus the density of the phases are required. thus the present solver provides the saturation index, rather than mass
or volume of solid. In future releases we hope to address this issue.

In addition, the quantity, N, of a species has different unit conventions
depending on the context. All aqueous species throughout the mathematical
formulation are given in mole/m$^3$. Surface species however, have units of
mole/m$^3$ in the composition matrix, whereas in the reaction matrix the mole
fraction convention is adopted. This is due to complications in defining the
reaction constant for reactions with multi-dentate species \cite{wang2013}{}.

\section{Newton solver strategies}

\subsection{A log-log formulation of the residual equations}

Logarithm transformations of the concentrations and residual equations are
certainly commonly used in Newton solver for chemical equations, see for example
the recent paper \cite{luo2015robust}. For the concentrations, which are the
unknowns in the system, using their logarithms as primary variables are the clear
advantage of imposing in a soft way the positivity of the concentrations. For the
residual equations, it seems natural to apply the logarithm to the reaction
equations as they get linear, see \eqref{eq:chemeqnice1}. However, to take the
logarithm of the constitutive equations, that is replace in the Newton algorithm
$[V^tx]_i = X_i$ with $\ln(V_i^tX)=\ln(X_i)$, seems less obvious. For example, the
authors in \cite{luo2015robust} do not mention this choice. In
\cite{wallgreening86}, the authors, motivated by results from geometric
programming, advocate for such treatment of the residual equations.

We use the same notations as in the first section. Assuming that all components
have activities equal to their concentrations, the chemical equilibrium reactions
are given by
\begin{equation}
  \label{eq:chemrec}
  \prod_{j=1}^{n_c} x_j^{a_{k,j}} = K_k,
\end{equation}
for $k=1,\ldots, n_e$. We follow the approach of \cite{wallgreening86} and do not
assume that the coefficients in $V$ are positive. We rewrite the conservation
equations given by \eqref{eq:conseq} as
\begin{equation}
  \label{eq:consv}
  \sum_{j=1}^{n_c} V_{j,k}^+ x_j -   \sum_{j=1}^{n_c} V_{j,k}^- x_j  =   M_k 
\end{equation}
where $V_{j,k}^{\pm}$ denotes the positive and negative par of $V_{j,k}$, so that
$V_{j,k} = V_{j,k}^+ - V_{j,k}^-$. Here $M_k$ is a given constant. We decompose
$M_k$ as $M_k=M_k^+ - M_k^-$, where $M_k^+=\max(0,M_k)$ and $M_k^-=-\min(0,M_k)$
so that either $M_k^+$ or $M_k^-$ are in fact zero. We rewrite these two sets of
equations in the synthetic form
\begin{equation}
  \label{eq:synthetic}
  f_i(x) = 1.
\end{equation}
In particular for $k = 1,\ldots, n_c - n_e$, we have
\begin{equation}
  \label{eq:deffi}
  f_{k + n_c - n_e}(x) = \frac{\sum_{j=1}^{n_c} (V_{j,k}^+ x_j) + M_k^-}{\sum_{j=1}^{n_c} (V_{j,k}^- x_j) + M_k^+} 
\end{equation}
The geometric programming approach consists of introducing the logarithm variables $\hat x_j =
\ln(x_j)$ and to solve, using Newton iterations, the residual equations
\begin{equation}
  \label{eq:logform}
  \ln(f_i(x)) = 0,
\end{equation}
i.e., $\ln(f_i(e^{\hat x_1}, \ldots, e^{\hat x_n})) = 0$. Let us briefly the
approach of \cite{wallgreening86}. We compute the derivative of $\ln f_{k + n_c - n_e}$
with respect to $\hat x_i$ and we obtain
\begin{align}
  \notag\fracpar{}{\hat x_i}\ln f_{k + n_c - n_e}  & = \fracpar{}{\hat x_i}\ln\big(\sum_{j=1}^{n}V_{j,k}^+x_j + M_k^-\big)  - \fracpar{}{\hat x_i}\ln\big(\sum_{j=1}^{n}V_{j,k}^+x_j + M_k^-\big)\\
  \notag  & = (\frac{V_{i,k}^+}{\sum_{j=1}^{n}V_{j,k}^+x_j + M_k^-} - \frac{V_{i,k}^-}{\sum_{j=1}^{n}V_{j,k}^-x_j + M_k^+})\frac{dx_i}{d\hat x_i}\\
\label{eq:defWik} & = \frac{V_{i,k}^+x_i}{\sum_{j=1}^{n}V_{j,k}^+x_j + M_k^-} -
\frac{V_{i,k}^-x_i}{\sum_{j=1}^{n}V_{j,k}^-x_j + M_k^+}
\end{align}
because $\frac{dx_i}{d\hat x_i} = x_i$. We denote by $W_{j,k}$ the right-hand side
in \eqref{eq:defWik}. It turns out that $f_{k + n_c - n_e}$ can be rewritten as
\begin{equation}
  \label{eq:geomprog}
  f_{k + n_c - n_e}  =D_k\prod_{W_{j,k}\neq 0 }\left(\frac{V_{j,k} x_j}{W_{j,k}}\right)^{W_{j,k}}
\end{equation}
where the constant $D_k$ is defined as 
\begin{equation*}
  D_k =
  \begin{cases}
    \left(\frac{M_k^+}{W_k^+}\right)^{W_k^+}&\text{ if $M_k^+>0$,},\\
    \left(\frac{M_k^-}{W_k^-}\right)^{W_k^-}&\text{ if $M_k^->0$}.
  \end{cases}
\end{equation*}
The residual equations as given by \eqref{eq:geomprog} take the form of a
geometric programming problem. In such case, explicit expression for the inverse
of the Jacobian are available. We refer to \cite{wallgreening86} for more details
on this aspect which is not our prime interest.

We are more interested in the efficiency of the algorithm, which is based on the
convexity of the residual equations. In \cite{wallgreening86}, the authors even
claim that, because of the convexity of every residual equation, the algorithm is
unconditionally convergent. We are unsure about this claim and have not found in
the literature evidence supporting it, but convexity of the residual equations can
be expected to bring stability. When the coefficients of $V$ are positive, the
residual equation given in \eqref{eq:deffi} takes the form
\begin{equation}
  \label{eq:linearfunc}
  g(x) = \sum_{j=1}^{n} \alpha_j x_j,
\end{equation}
for positive coefficients $\alpha_j$. Function of the form above retain their
convexity when changing to logarithmic variables. We detail the proof since it is
not done in \cite{wallgreening86}. The Hessian of $\ln(g)$ with respect to $\hat
x$, which we denote $H$, is given by
\begin{equation*}
  H_{l,j} = \frac{\partial}{\partial \hat x_l\partial \hat x_i}\ln(g) = 
  \begin{cases}
    -g^{-2}\alpha_ix_i\alpha_lx_l  & \text{ if }i\neq l,\\
    -g^{-2}\alpha_i^2x_i^2 + g^{-1}\alpha_ix_i&\text{ if }i=l.
  \end{cases}
\end{equation*}
Hence, for any $u\in\Real^{n}$, we have
\begin{equation*}
  u^t H u = -\abs{v}^2 + u\cdot v
\end{equation*}
for $v\in\Real^n$ defined as
\begin{equation*}
  v_i = \frac{\alpha_ix_iu_i}{g}.
\end{equation*}
It follows that $u_i -v_i = (\frac{g}{\alpha_ix_i} - 1)v_i $ and therefore
\begin{equation*}
  u^t H u = \sum_{i=1}^n (\frac{\sum_{j=1}^n\alpha_jx_j}{\alpha_ix_i} - 1)v_i^2  = \sum_{i=1}^n\sum_{\substack{j\neq i,\\j=1}}^n\frac{\alpha_jx_j\alpha_ix_iu_i^2}{g^2}\geq 0,
\end{equation*}
so that $\ln(g)$ is convex.

\subsection{Apriori bounds based on conservation of constitutive elements}

Let us denote by $X\in\Real^{n_c - n_e}$ the concentration of each of the
constitutive elements. When $V$ is given by the composition matrix, the vector $X$
corresponds to $M$ in \eqref{eq:defMN}. Thus, using the notations of the previous
section, we have
\begin{equation*}
  V^t x = X.
\end{equation*}
The coefficients of $V$ and $x$ are positive.  Therefore, for each specie
$i=\{1,\ldots,n_c\}$, we get
\begin{equation*}
  x_i V_{ij} \leq X_j,
\end{equation*}
for all $j\in \{1,\ldots,n_c - n_e\}$. Given the total concentrations $X_j$, we
end up with the following upper bound for $x_i$,
\begin{equation}
  \label{eq:upperbound}
  x_i \leq \min_{j\in \{1,\ldots,n_c - n_e\}} \frac{X_j}{V_{ij}} = m_i.
\end{equation}
This simple bound turns out to improve significantly the robustness of the method,
as shown in particular in the first numerical test below. They are used for
chopping: If, after a Newton update, $x_i>m_i$, then we set $x_i = m_i$.

\section{Transport Equations}

We consider the general case introduced in the first section but now the chemical
species are transported. We denote by $u_i$ the flux of each component
species. The governing equations are then
\begin{equation}
  \label{eq:transgen}
  \fracpar{N_i}{t} + \dive(u_i) = F_i(N)
\end{equation}
where the source term $F_i$ is defined in \eqref{eq:defFi}. Using that $V^tF = 0$, we obtain two set
of equations from \eqref{eq:transgen} 
\begin{equation}
  \label{eq:transcons}
  \fracpar{M}{t} + \dive(V^tu) = 0
\end{equation}
and
\begin{equation}
  \label{eq:transnoncons}
  \fracpar{P}{t} + \dive(W^tu) = W^tAG
\end{equation}
where $M$ and $P$ are defined in \eqref{eq:defMN}. If we assume that the time
scale for the chemical equations is much faster than the time scale for transport
in \eqref{eq:transnoncons}, then we obtain $W^tA^tG=0$, which is equivalent to
$G=0$. The governing equations are therefore
\begin{subequations}
  \label{eq:govtransport}
  \begin{align}
    \fracpar{X}{t} + \dive(V^tu(x)) &= 0,\\
    G(x)                            &= 0,\\
    V^t x                           &= X.
  \end{align}
\end{subequations}
The flux $u_i(x)$ of each species is in the case of a single fluid phase given by
\begin{equation}
  \label{eq:deffluxes}
  u_i(x) =
  \begin{cases}
    x_i U &\text{ if the component $i$ is a dissolved component,}\\
    0 &\text{ if the component $i$ belongs to the solid phase.}
  \end{cases}
\end{equation}
Above $U$ denotes the fluid phase velocity. For a porous media with permeability
$K$ and an incompressible fluid, $U$ satisfies $\dive U=0$ and $U=-K\grad p$. 

\section{Test cases}

\subsection{Alkalinity-pH equation}

To analyze the performance of the chemical solver, we use the test case presented
in \cite{munhoven2013mathematics}. The aim is to compute the composition of sea
water for a given alkalinity. The method used by Munhoven is very different from
ours: Instead of solving directly the full system of equations, the system is
reduced analytically to a scalar equation with one unknown, the \ch{H^+}
concentration. Then, the equation takes the form of a third order polynomial and
robust algorithms to solve this polynomial are derived, which also include apriori
bounds for the roots. The resulting algorithms are very effective. However, the
method is inherently tailored to this particular chemical system and activities
are not included. The system includes 18 chemical species given by
\begin{align*}
  \label{eq:labellistakl}
  &\ch{H2CO3},\  \ch{HCO3-},\  \ch{CO3^{2-}},\  &&
  \ch{H2O},\  \ch{H^{+}},\  \ch{OH^{-}},\\
  &\ch{B(OH)_4-},\  \ch{B(OH)_3},\ &&
  \ch{H3PO4},\  \ch{H2PO4^{-}},\  \ch{HPO4^{2-}},\  \ch{PO4^{3-}},\ \\
  &\ch{H4SiO4},\  \ch{H3SiO4-},\ &&
  \ch{HSO4-},\  \ch{SO4^{2-}},\ \\
  &\ch{HF},\  \ch{F-}.
\end{align*}
They are involved in the following chemical equations,
\begin{subequations}
  \label{eq:reactionsalk}
  \begin{align}
    &\ch{H2CO3 <->[ $K_{C1}$ ] H^{+} + HCO3^-},\\
    &\ch{HCO3- <->[ $K_{C2}$ ] H^{+} + CO3^{2-}},\\
    &\ch{H2O <->[ $K_{w}$ ] H^+ + OH^-},\\
    &\ch{B(OH)4^- <->[ $K_{B}$ ] B(OH)3 + OH^-},\\
    &\ch{H3PO4 <->[ $K_{P1}$ ] H^{+} + H2PO4-},\\
    &\ch{H2PO4^{-} <->[ $K_{P2}$ ] H^{+} + HPO4^{2-}},\\
    &\ch{HPO4^{2-} <->[ $K_{P3}$ ] H^{+} + PO4^{3-}},\\
    &\ch{H4SiO4 <->[ $K_{Si}$ ]  H^{+} +  H3SiO4-},\\
    &\ch{HSO4- <->[ $K_{SO4}$ ] H^{+} + SO4^{2-}},\\
    &\ch{HF <->[ $K_{F}$ ] H^{+} + F-}.
  \end{align}
\end{subequations}
The total alkalinity is given in this case by
\begin{multline}
  \label{eq:alkdef}
  \ch{Alk_T} = \ch{[HCO3^-]} + 2\ch{[CO3^{2-}]}\\ + \ch{[B(OH)4-]} + \ch{[OH^{-}]} +
  \ch{[HPO4^{2-}]} + 2\ch{[PO4^{3-}]} + \ch{[H3SiO4-]} \\ - \ch{[H^+]} -
  \ch{[HSO4-]} - \ch{[HF]} - \ch{[H3PO4]}.
\end{multline}
We use the routines provided in \texttt{SolveSAPHE}, which is available in the
supplementary material of \cite{munhoven2013mathematics}, to compute the chemical
kinetic constants at temperature $T= \SI{275.15}{\kelvin}$, pressure
$p=\SI{0}{\bar}$ and salinity coefficient $s = 35$. We have
\begin{center}
\def\arraystretch{1.4}
\begin{tabular}[h]{rrrrrrrrrr}
  $pK_{C_1}$&$pK_{C_2}$&$pK_B$&$pK_{P_1}$&$pK_{P_2}$&$pK_{P_3}$&$pK_{Si}$&$pK_{SO4}$&$pK_{F}$&$pK_w$\\
  \hline
  6.1&9.3&8.9&1.6&6.2&9.3&9.8&0.58&2.5&14
\end{tabular}
\end{center}

\begin{figure}[h]
  \centering
  \includegraphics[width=0.5\textwidth]{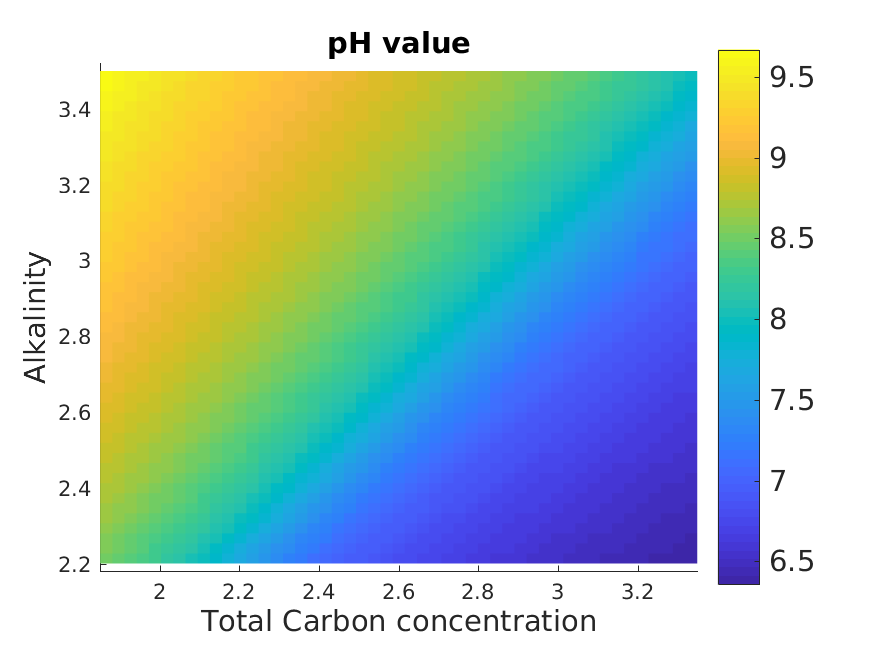}
  \caption{plot of pH values}
  \label{fig:phplot}
\end{figure}

The constant inputs are given by
\begin{center}
\def\arraystretch{1.4}
\begin{tabular}[h]{cccccc}
  &\ch{[F]_T}&\ch{[SO4]_T}&\ch{[B]_T}&\ch{[P]_T}&\ch{[SiO4]_T}\\
  \hline
  \si{\mol\per\litre}&0.068&28&0.42&5e-7&5e-6
\end{tabular}
\end{center}
Note that the total concentrations of \ch{P} and \ch{Si} that are given in
\cite{munhoven2013mathematics} differ from the one from the code which is provided
in the supplementary online material. We consider the values given in the code
which are consistent with the results that are shown in the paper. We look at the
test case SW2 from \cite{munhoven2013mathematics}. We compute the chemical
composition of the mixture for values of \ch{[C]_T} ranging from \num{1.85} to
\SI{3.35}{\milli\mol\per\kilogram} and values of $\ch{[Alk_{T}]}$ ranging from \num{2.2}
to \SI{3.5}{\milli\mol\per\kilogram}. We use a resolution of $50\times 50$. The
values for the pH are given in Figure \ref{fig:phplot} and they correspond to the
values obtained in \cite{munhoven2013mathematics}.

In Figure \ref{fig:numiter}, we plot the number of Newton iterations necessary for
convergence. The convergence criteria in this case is $\abs{\ch{[H^{+}]}^{n+1}
  -\ch{[H^{+}]}^{n}}/\ch{[H^{+}]}^{n}<\epsi$, and we use $\epsi=\num{1e-8}$, as in
\cite{munhoven2013mathematics}. We consider two cases for different choices of the
initial guess. In the first case, we use an educated initial guess, based on the
physical bounds on the unknowns. In the second case, the initial guess is set
uniformly to \SI{1}{\mol\per\litre} for all the unknown. In both cases, the method
always converges and the second case uses only a few extra iterations more than
the first case, showing the robustness of the approach.

\begin{figure}[h]
  \begin{center}
    \begin{tabular}[c]{cc}
    \includegraphics[width=0.5\textwidth]{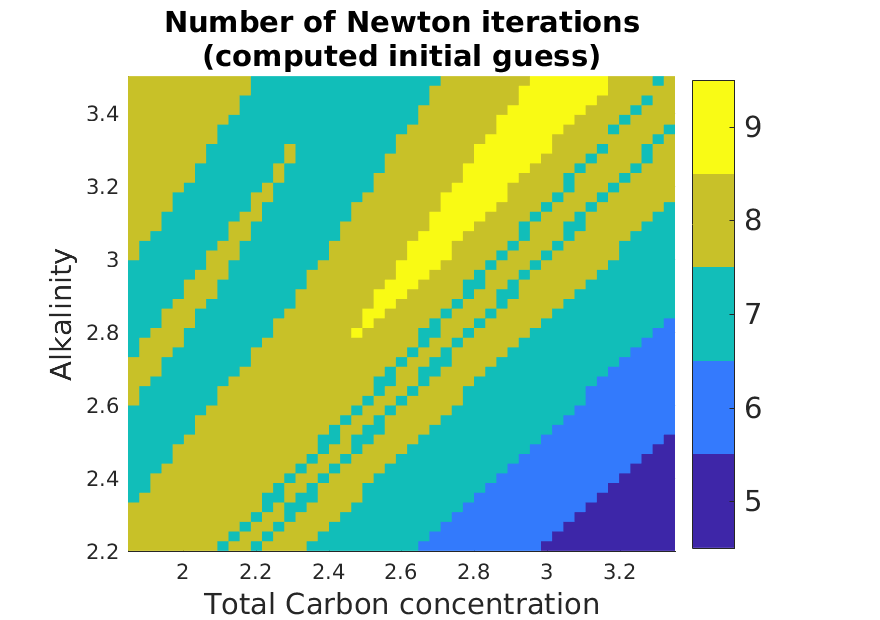}&
    \includegraphics[width=0.5\textwidth]{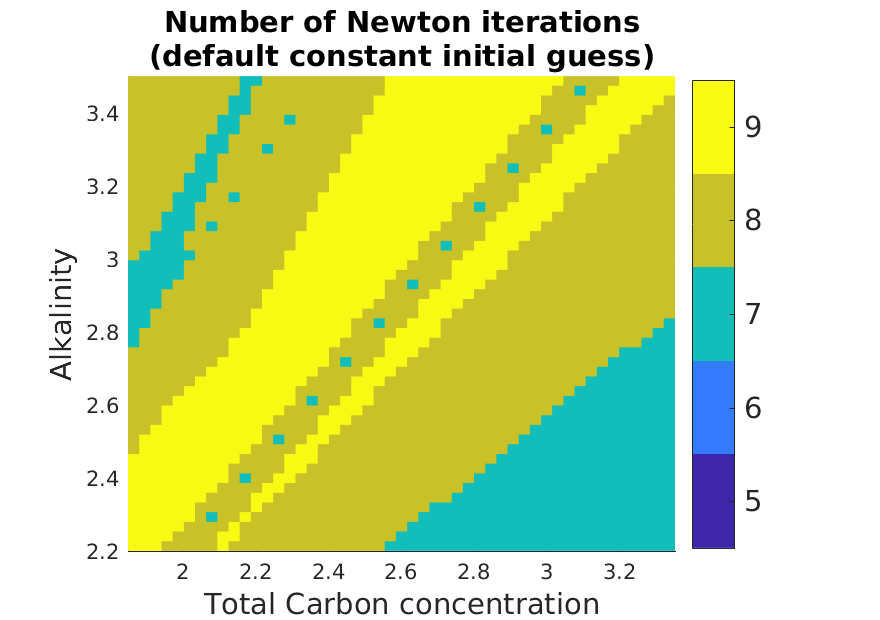}
  \end{tabular}
  \end{center}
  \caption{Number of Newton iterations. We observe that the number of iterations
    is smaller for the computed initial guess. However, even for rough uniform
    initial guess, the method always converges.}
  \label{fig:numiter}
\end{figure}

We use this test case to assess the importance of the physical bound in the
convergence of the Newton algorithm. To do so, we switch off the chopping of the
variables after each Newton updates. The results are presented in Figure
\ref{fig:numiter}. The method is not always convergent and require otherwise
significantly more Newton steps. This result shows the important of the chopping
step.

\begin{figure}[h]
  \begin{center}
    \includegraphics[width=0.5\textwidth]{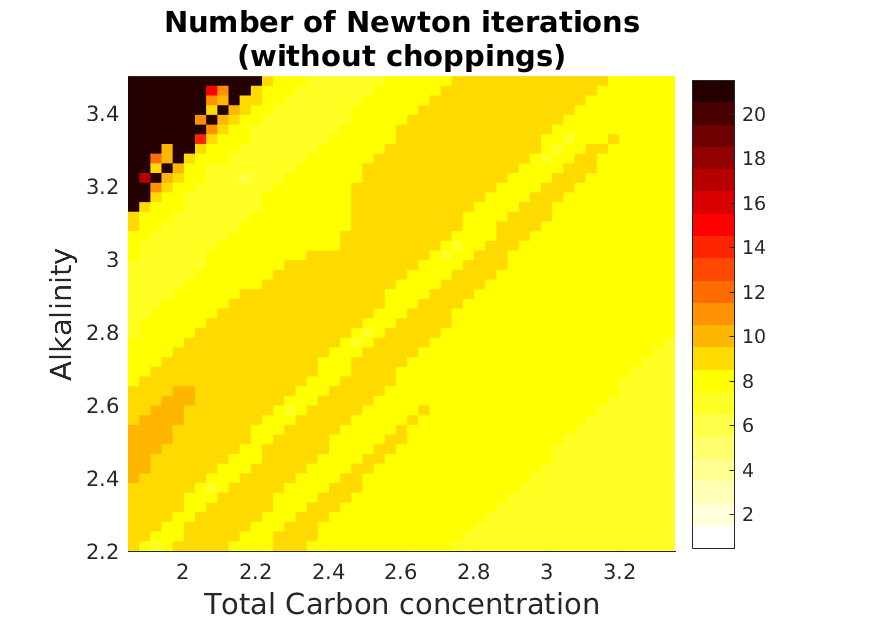}
  \end{center}
  \caption{Number of Newton iterations in the case where the chopping of the
    variables is not used. For some of the parameters, which corresponds to the
    value above 21 in the plot, the method does not converge.}
  \label{fig:numiter}
\end{figure}

\subsection{Equilibrium with an electrostatic surface}

We consider here an amphoteric surface with the sorption of \ch{H^+}, \ch{Na^+},
and \ch{Cl^-}. We have one group site which we denote $\ch{SO}$, being an
arbitrary hydroxide surface, which can evolve as the following species
\begin{equation*}
  \ch{>SOH},\   \ch{>SO^-},\  \ch{>SOH2^+},\  \ch{>SONa},\  \ch{>SOH2Cl},
\end{equation*}
depending on the sorbed component. We use the triple layer model to compute the
equilibrium of this system. We use the same notation as in the section above where
the triple layer model is introduced but, since there is only a single surface, we
drop the index $o$. Each surface component will contribute to the charge of a
given layer. In this example, \ch{H^+} sorbs directly to the mineral surface, and
is called an inner sphere sorption complex. The species \ch{Na^+} and \ch{Cl^-}
are geometrically limited, and can not sorb directly to this mineral surface. Thus
they occupy the second layer ($p=2$) and contribute their charge to $p=2$.  Therefore,
the charged components \ch{>SO^-} and \ch{>SOH2^+} contribute their charge to the
first layer. The non-charged component may contribute through their polarization
as they spread across the two first layer. In this case, \ch{>SoNa} contributes
with a negative charge layer to the first layer and a positive charge to the
second layer. For \ch{>SOH2Cl}, it is the opposite. The component \ch{>SOH} is
inactive in this context, see Figure \ref{fig:surfchem} for a table overview.

The unknowns are the species concentrations,
\begin{equation}
  \label{eq:surfspec}
  \text{\begin{tabular}[c]{*{10}{c@{,\hspace*{2mm} }}c}
      \ch{>SOH}&   \ch{>SO^-}&  \ch{>SOH2^+}&  \ch{>SONa}&  \ch{>SOH2Cl} & \ch{H^+} & \ch{OH^-} & \ch{H2O} & \ch{Na^+} & \ch{Cl^-} & \ch{NaCl}
    \end{tabular}}
\end{equation}
and the corresponding activities, the potentials and charges of the layers
($\Psi_i$, $\sigma_i$, for $i=1,2,3$) and the ionic strength $I$. We index the
species using the same ordering as in \eqref{eq:surfspec}. We denote by $x_s$ the
concentration of the surface species ($x_{s,i}=x_i$ for $i=1,\ldots,5$) and $x_a$
for the aqueous species ($x_{a,i}=x_{i + 5}$ for $i=1,\ldots,6$). The total number
of unknowns is equal to 29.

Let us present the governing equations. The chemical equilibrium equations are
\begin{subequations}
  \label{eq:surfchemexgoveq}
\begin{equation}
  A\hat x + A\hat \gamma = \hat K,
\end{equation}
for $A\in\Real^{n_r\times n_c}$, see Figure \ref{fig:matricesurfchem}.  the definition
of the activities for the aqueous species follow from \eqref{eq:aqactivity},
\begin{equation}
  \hat\gamma_{\ch{NaCl}} = \hat\gamma_{\ch{H2O}} = 0\quad\text{ and }\quad \hat\gamma_{\ch{H^+}} = -\hat\gamma_{\ch{OH}} = \hat\gamma_{\ch{Na^+}} = - \hat\gamma_{\ch{Cl^-}} = f(I)
\end{equation} 
where the definition of the function $f$ can be inferred from
\eqref{eq:aqactivity}. The equation for the ionic strength is linear and given by
\eqref{eq:ionicstrengthdef}, which we rewrite as
\begin{equation}
  I = \kappa^tx_a
\end{equation}
for $\kappa=(1, -1, 0, 1, -1, 0)^t$ in this case. The equations for the activities
of the surface species are given by \eqref{eq:surfAct}, which we rewrite as
\begin{equation}
  \hat\gamma_s = c_1 B\Psi,
\end{equation}
for a matrix $B\in\Real^{n_s\times 3}$. Our goal here is to present the form of
the equations and to simplify the notations we denote generically by $c_i$ the
constant physical terms that enter the equations. The defining equations for the
charge are
\begin{equation}
  \sigma = c_2 B^tx_s + G(\sinh(\frac{F\Psi_3}{2RT})),
\end{equation}
where $G$ is a function in $\Real^3$ which is non-zero only for the third
component. The capacitance relations are
\begin{equation}
  \sigma_1 = C_1(\Psi_1 - \Psi_2),\quad \sigma_2 = C_2(\Psi_3 - \Psi_2).
\end{equation}
The charge balance is
\begin{equation}
  \sigma_1 +   \sigma_2 +  \sigma_3 = 0.
\end{equation}
\end{subequations}
The governing equations are given by \eqref{eq:surfchemexgoveq} and form a system
of 24 equations. We have 5 master components, given by
\begin{center}
\begin{tabular}[c]{*{5}{c@{,\quad }}}
  \ch{H}&   \ch{O}&  \ch{Na}&  \ch{Cl}&  \ch{>SO}
\end{tabular}
\end{center}
so that we match the 29 unknowns. Let us now set a numerical test. We vary the
total hydrogen concentration $\ch{[H]}_T$ from \num{1e-4} to
\SI{1e-10}{\mol\per\litre}. For the master component \ch{Na} and \ch{Cl}, we use
constant total concentrations given by
$\ch{[Na]}_T=\ch{[Cl]}_T=\SI{1e-2}{\mol\per\litre}$. The normalized water
concentration is also kept constant and equal to $\SI{1}{\mol\per\litre}$.

\begin{figure}[h]
  \centering
\def\arraystretch{1.4}
  \begin{tabular}[t]{l|r}
    reaction                        & equilibrium constant \\
    \hline
    \ch{>SOH <> >SO^- + H^+},       & $10^{-7.5}$ \si{\mol\per\litre}\\
    \ch{>SOH + H^+ <> >SOH2^+},     & $10^{2}$ \si{(\mol\per\litre)^{-1}}\\
    \ch{>SO^- + Na^+ <> >SONa},     & $10^{-1.9}$ \si{(\mol\per\litre)^{-1}}\\
    \ch{>SOH2^+ + Cl^- <> >SOH2Cl}, & $10^{1}$ \si{(\mol\per\litre)^{-1}}\\
    \ch{H2O  <> H^+  + OH^-},       & $10^{-14}$ \si{\mol\per\litre}\\
    \ch{NaCl <> Na^+ + Cl^-},       & $10$ \si{\mol\per\litre}
  \end{tabular}
\hspace*{2cm}
\begin{tabular}[t]{r@{: }cc}
  \multicolumn{1}{r}{}& layer 1 & layer2 \\
  \hline
  \ch{>SOH}    & 0       & 0   \\
  \ch{>SO^-}   &-1       & 0   \\
  \ch{>SOH2^+} & 1       & 0   \\
  \ch{>SONa}   & -1      & 1   \\
  \ch{>SOH2Cl} & 1       & -1   \\
\end{tabular}
\caption{Reaction and surface species contribution}
\label{fig:surfchem}
\end{figure}

\newcolumntype{C}{@{\hspace*{1pt}}>{\begin{center}\arraybackslash}m{4mm}<{\end{center}}@{\hspace*{1pt}}}
\begin{figure}[h]
  \centering
  \def\arraystretch{0.1}
  $A=
  \left(\text{
      \begin{tabular}[c]{*{11}{C}}
        1  &1 &0  &0  &0  &-1 &0  &0  &0  &0 &0\\[-3mm]
        0  &0 &1  &1  &-1 &0  &0  &0  &0  &0 &0\\[-3mm]
        1  &0 &0  &0  &0  &0  &1  &-1 &0  &0 &0\\[-3mm]
        -1 &0 &0  &0  &0  &0  &0  &-1 &1  &0 &0\\[-3mm]
        0  &0 &-1 &0  &0  &0  &-1 &0  &0  &1 &0\\[-3mm]
        0  &0 &0  &-1 &0  &0  &0  &0  &-1 &0 &1\\[-5mm]
      \end{tabular}}\right)$
  \hspace*{2cm}
  $B^t=
  \left(\text{
      \begin{tabular}[c]{*{3}{C}}
        0  &0  &0 \\[-3mm]
        -1 &0  &0 \\[-3mm]
        1  &0  &0 \\[-3mm]
        -1 &1  &0 \\[-3mm]
        1  &-1 &0 \\[-5mm]
      \end{tabular}}\right)$
  \caption{Matrices involved in this test case}
  \label{fig:matricesurfchem}
\end{figure}

\begin{figure}[h]
  \centering
  \begin{tabular}[c]{cc}
  \includegraphics[width=0.5\textwidth]{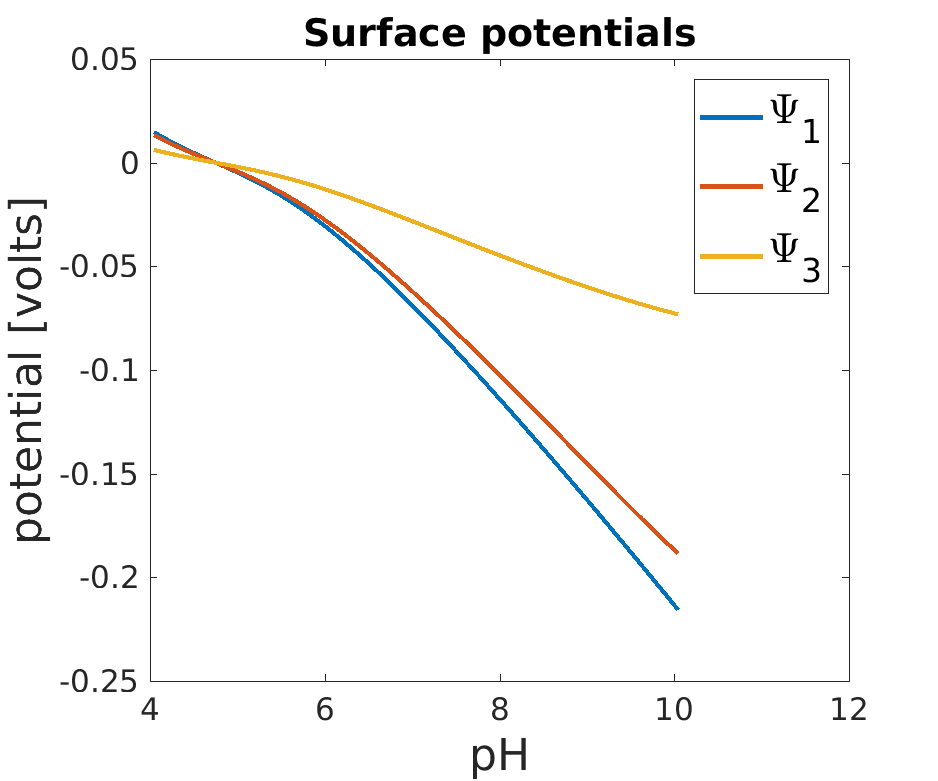}
  \includegraphics[width=0.5\textwidth]{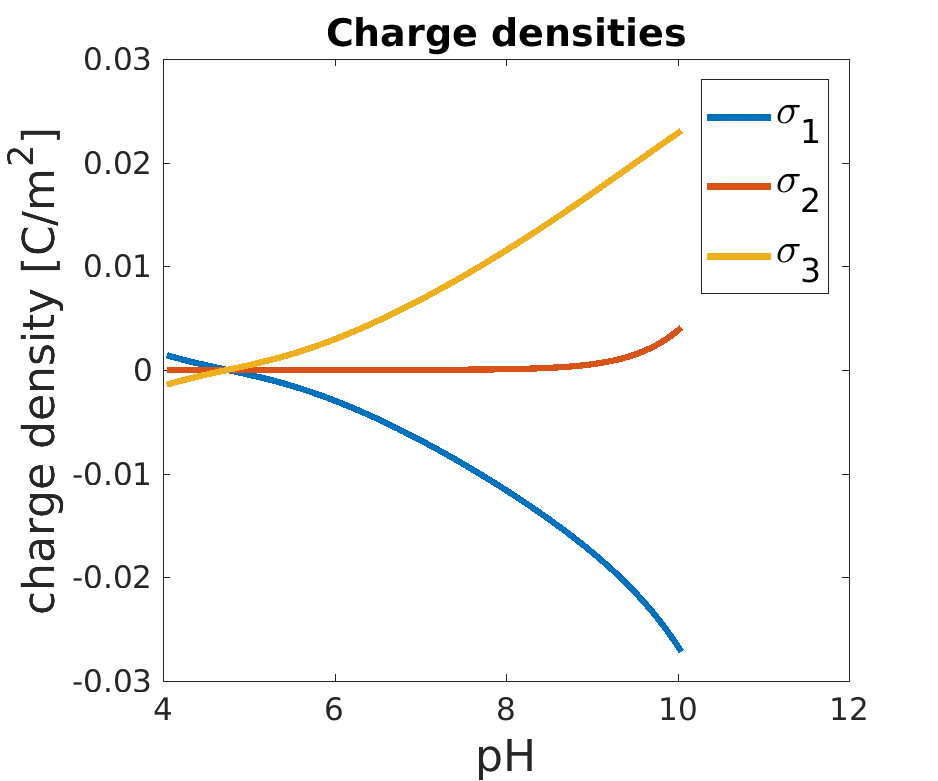}
\end{tabular}
\caption{Plots of the surface potential and charge densities as a function of the
  solution pH}
  \label{fig:surfchemres}
\end{figure}

The potential and charge of each layer of the chemical system are plotted in
figure \ref{fig:surfchemres}. It can be seen that these quantities vary smoothly
as a function of pH, which is expected. Further, the charge of each layer sums to
zero, as is enforced in the model. Finally, it can be seen that near pH=3 the
potential and charge of all layers is zero. This is known as the point of zero
charge and is a defining characteristic of an amphoteric surface.

\clearpage

\section*{Acknowledgment}
X. Raynaud thanks support from the Norwegian Research Council (KPN 280651). This
work is funded in part by the Center for Frontiers of Subsurface Energy Security,
an Energy Frontier Research Center funded by the U.S. Department of Energy, Office
of Science, Basic Energy Sciences under Award \#DE-SC0001114.

\end{document}